\documentclass[10pt,conference,onecolumn]{IEEEtran}

\pdfoutput=1

\usepackage{amssymb,amsmath} 

\usepackage{cite}

\ifCLASSINFOpdf
  \usepackage[pdftex]{graphicx}
 
  \DeclareGraphicsExtensions{.pdf,.jpg,.png}
\else
 
\fi

\usepackage{enumerate}
\usepackage{subfig}
\usepackage{multicol}
\usepackage{float}
\usepackage{mdwtab}

\usepackage{url}

\usepackage[utf8]{inputenc}

\usepackage[table,xcdraw]{xcolor}

\begin{document}

%
\title{Maintenance of Automated Test Suites in Industry: An Empirical study on Visual GUI Testing}


\author{Emil Al\'egroth\\
Chalmers University of Technology\\ Department of Computer Science \\and Engineering \\ SE-412 96 G\"oteborg, Sweden\\ emil.alegroth@chalmers.se
\and
Robert Feldt\\
Software Engineering Research Lab\\ School of Computing\\ Blekinge Institute of Technology\\ SE-371 79 Karlskrona\\Sweden\\ Robert.Feldt@gmail.com
\and
Pirjo Kolstr\"om\\
Saab Sensis ATM Sweden\\ Ljungadalsgatan 2\\ 352 46 V\"axj\"o\\ Sweden\\ Pirjo.Kolstrom@Saabgroup.com
}

\maketitle

\begin{abstract}
\textbf{Context:}  Verification and validation (V\&V) activities make up 20 to 50 percent of the total development costs of a software system in practice.
Test automation is proposed to lower these V\&V costs but available research only provides limited empirical data from industrial practice about the maintenance costs of automated tests and what factors affect these costs. 
In particular, these costs and factors are unknown for automated GUI-based testing.

\textbf{Objective:} This paper addresses this lack of knowledge through analysis of the costs and factors associated with the maintenance of automated GUI-based tests in industrial practice.

\textbf{Method:} An empirical study at two companies, Siemens and Saab, is reported where interviews about, and empirical work with, Visual GUI Testing is performed to acquire data about the technique's maintenance costs and feasibility.

\textbf{Results:} 13 factors are observed that affect maintenance, e.g. tester knowledge/experience and test case complexity.
Further, statistical analysis shows that developing new test scripts is costlier than maintenance but also that frequent maintenance is less costly than infrequent, big bang maintenance.
In addition a cost model, based on previous work, is presented that estimates the time to positive return on investment (ROI) of test automation compared to manual testing. 

\textbf{Conclusions:} It is concluded that test automation can lower overall software development costs of a project whilst also having positive effects on software quality. 
However, maintenance costs can still be considerable and the less time a company currently spends on manual testing, the more time is required before positive, economic, ROI is reached after automation.
\end{abstract}


\IEEEpeerreviewmaketitle

\section{Introduction} \label{intro}
The cost of testing is a key challenge in the software industry, reported by both academia and industry to be sometimes upwards of 50 percent of total development cost and rarely below 20 percent~\cite{ellims2006economics, hailpern2002software,ericson1997tim}. 
In addition, software industry is moving towards a faster and more agile environment with emphasis on continuous integration, development and deployment to customers~\cite{olsson2012climbing}.
This environment puts new requirements on the speed of testing and presents a need for quicker and more frequent feedback on software quality. Often this is used as an argument for more test automation.

Many test automation techniques have been proposed, such as automated unit testing~\cite{olan2003unit, gamma1999junit, ellims2006economics} , property-/widget-based graphical user interface (GUI) testing~\cite{holmes2006automating, hackner2008test, vizulisself}, and Visual GUI Testing (VGT)~\cite{borjesson2012vgt, alegroth2013vgt, alegrothvisual}.
However, even though empirical support exists for the techniques use in practice~\cite{holmes2006automating, hackner2008test, vizulisself, alegroth2013vgt, leotta2014visual}, less information has been provided on the costs associated with test automation.
Further, even if theoretical cost models have been presented, there is a lack of models grounded in actual, empirical data from industrial software systems.
Related work~\cite{rafi2012benefits, wagner2006model, karhu2009empirical, liu2000platform, berner2005observations, fewster1999software, sjosten2006costs, leotta2013capture}, has also reported on what factors affect the maintenance of automated testing but not explicitly what factors affect automated GUI based testing.

In this work we address these gaps in knowledge through an embedded empirical study~\cite{runeson2009guidelines} with the goal to identify what costs are associated with automated GUI-based testing, here represented by VGT, in industrial practice.
First, an interview study at Siemens provides qualitative information regarding the usage and maintenance of VGT from a longer project perspective (Seven months).
Second, an empirical study at Saab that in detail identifies the costs of maintaining VGT suites at different levels of degradation.
A heavily degraded test suite is maintained for an industrial system to acquire (worst case) cost information.
The maintained suite is then migrated (maintained) to another variant of the industrial system to acquire information about frequent test maintenance (best case).
The study results show that the frequency of maintenance affects the maintenance cost but also that maintenance is less costly than development of the scripts.
Statistical analysis on a finer level of script granularity also showed that maintenance of the GUI components the scripts interact with is less costly than maintenance of the test case scenario logic.
We also present a correlation analysis that shows that the changes to the manual test cases is a poor estimator for the maintenance costs of the automated test cases.

In addition, observations made during the study support related work~\cite{karhu2009empirical,berner2005observations} that there are several factors, both technical and context dependent, which influence the maintenance.
In total, thirteen (13) factors are reported and discussed in terms of their impact on automated GUI based testing. 
Further, the acquired quantitative metrics are modeled using a theoretical cost model defined in previous work~\cite{alegrothvisual, berner2005observations}.
The model depicts the time spent on VGT maintenance, best and worst case, to be compared to the cost of manual testing at Saab as well as a fictional, but realistic, context where 20 percent of the project development time is spent on testing.

From the study we conclude that VGT maintenance provides positive return on investment in industrial practice but is still associated with significant maintenance cost that should not be underestimated.
As such this paper provides several contributions to the body of knowledge on automated testing regarding maintenance of GUI-based testing, i.e. explicit cost data from practice as well as factors that influence the size of these costs.

The continuation of this manuscript is structured as follows.
Section \ref{rw} will present related work, followed by Section \ref{method} that will present the research methodology.
The paper continues by presenting the acquired results in Section \ref{res}, which are then discussed in Section \ref{disc}.
Finally the paper is concluded in Section \ref{conc}.

\section{Related work} \label{rw} 
Manual software testing is associated with problems in practice such as high cost and tediousness and error-proneness~\cite{grechanik2009maintaining, grechanik2009creating, finsterwalder2001automating, leitner2007reconciling, memon2002gui, dustin1999automated}.
Despite these problems, manual testing is still extensively used for system and acceptance testing in industrial practice.
One reason is because state-of-practice test automation techniques primarily perform testing on lower levels of system abstraction, e.g. unit testing with JUnit~\cite{cheon2006simple}.
Attempts to apply the low level techniques for high level testing, e.g. system and acceptance tests, have resulted in complex test cases that are costly and difficult to maintain, presenting a need for high level test automation techniques~\cite{autorev2012, gutierrez2006generation, berner2005observations, grechanik2009creating, horowitz1993g, sjosten2006costs, zaraket2012guicop, finsterwalder2001automating}.
Another reason for the lack of automation is presented in research as the inability to automate all testing~\cite{autorev2012,berner2005observations, rafi2012benefits, afzalexperiment, itkonen2007defect}.
This inability comes from the inability of scripted test cases to identify defects that are not explicitly asserted, which infers a need for, at least some level of, manual or exploratory testing~\cite{Itkonen2005}.

GUI-level testing can be divided into three chronologically defined generations.
The first and second generation techniques are performed either by capturing the exact coordinates (first generation) where a user interacts with the system under test (SUT) on the screen or through the GUI components' properties (second generation)~\cite{adamoliautomated, andersson2004video, memon2002gui}.
However, first generation test scripts are sensitive to change of the SUT's GUI layout, even minor changes to the GUI can make entire test suites inoperable, leading to high maintenance costs.
Second generation scripts are more robust and therefore used in industrial practice, e.g. with tools like Selenium~\cite{holmes2006automating} or QTP. 
However, the technique is still sensitive to changes to GUI components and only has limited support for automation of distributed systems and systems built from custom components~\cite{sjosten2006costs}.
These limitations originate in the technique's approach of interacting and asserting the GUI model rather than the GUI shown to the user on the computer monitor.
This approach requires the tools' to have access to the SUT's GUI library or other hooks into the SUT, which limits the tools use for SUT's written in certain programming languages and/or certain GUI libraries.
Consequently, second generation tools are considered robust but unflexible, leaving a need for a more flexible automated test approach.
However, because of the robustness and access to GUI state information, current research into GUI-based testing has primarily focused on this approach for a wide range of application areas, including model-based GUI testing, desktop application testing, etc.~\cite{gao2015, vos2015testar}.

The third generation, also referred to as Visual GUI Testing (VGT)~\cite{borjesson2012vgt, alegroth2013vgt, alegrothvisual}, instead uses image recognition that allows VGT tools, e.g. Sikuli~\cite{yeh2009sikuli} or JAutomate~\cite{alegroth2013jautomate}, to interact with any GUI component shown to the user on the computer monitor.
As a consequence, VGT has a high degree of flexibility and can be used on any system regardless of programming language or even platform.
Combined with scenario-based scripts, the image recognition allows the user to write testware applications that can emulate human user interaction with the SUT.
Previous research has shown that VGT is applicable in practice for the automation of manual system tests~\cite{borjesson2012vgt, alegroth2013vgt, alegrothvisual}.
However, only limited information has been acquired regarding the maintenance costs associated with the technique~\cite{alegroth2013vgt, alegrothvisual,leotta2014visual}.

Test maintenance is often mentioned in related work on automated testing but empirical data on maintenance costs from real, industrial projects are limited. 
There is also a lack of cost models based on such data and what factors affect the maintenance costs.
In their systematic review on the benefits and limitations of automated testing, Rafi et al. only identified four papers that presented test maintenance costs as a problem~\cite{rafi2012benefits}, yet only one of these papers addressed cost and then in the form of theoretical models~\cite{berner2005observations}.
Empirical papers exits that report maintenance costs but for open source software, e.g. ~\cite{leotta2013capture, leotta2014visual, NguyenASE2013}, but the number of papers with maintenance costs of automated test techniques on industrial systems are limited, e.g.~\cite{sjosten2006costs, borjesson2012vgt, alegroth2013vgt, alegrothvisual}.
Whilst our previous work reports that VGT can be applied in practice~\cite{borjesson2012vgt, alegroth2013vgt} but that there are challenges~\cite{alegrothvisual}, Sj{\"o}sten-Andersson and Pareto reports the problems of using second generation tests in practice~\cite{sjosten2006costs}.
Instead, maintenance is mostly discussed theoretically and presented through qualitative observations from industrial projects~\cite{karhu2009empirical, berner2005observations}.
For instance, Karhu et al.~\cite{karhu2009empirical} performed an empirical study where they observed factors that affect the use of test automation in practice, e.g. that maintenance costs must be taken into account and that human factors must be considered, but the paper does not  present any quantitative support for these observations.
The observations made by Karhu et al. are supported by Berner et al~\cite{berner2005observations} that also proposes a theoretical cost model for maintenance of automated testing, but yet again no empirical quantitative data is presented to support the qualitative observations.

There are many factors that affect the maintenance costs of automated tests, e.g. test design and strategies used for implementation.
As reported by Berner et al.~\cite{berner2005observations}, design of the test architecture is an important factor that is generally overlooked.
Another factor is the lack of architectural documentation of the testware and that few patterns or guidelines exist that promote the implementation of reusable and maintainable tests.
Additionally, many companies implement test automation with the wrong expectations and therefore abandon the automation, sometimes after considerable investment~\cite{karhu2009empirical,berner2005observations}.
However, not all factors that affect maintenance of automated testing are general and it is likely that not all factors have yet been identified.
As such, this work, in association with our previous work~\cite{alegrothvisual}, contribute to the knowledge about the factors that should be taken into account to lower maintenance costs.

\begin{figure}[t!]
  \centering
  \includegraphics[scale=0.5]{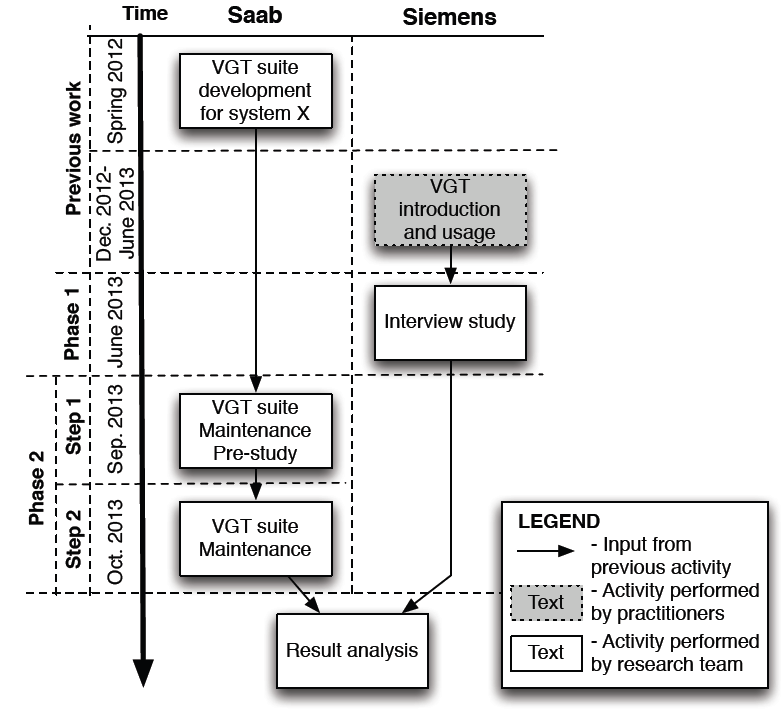}
  \caption{An overview of the methodology used during this work to acquire the study results.} 
  \label{meth}
\end{figure}

\section{Methodology} \label{method} 
The study's methodology was divided into two phases, as shown in Figure \ref{meth}. 
VGT maintenance was evaluated at two companies, which were chosen through convenient sampling due to the limited use of VGT in practice.
The two companies used different VGT tools, i.e. Sikuli~\cite{yeh2009sikuli} and JAutomate~\cite{alegroth2013jautomate}, but, as presented in previous work~\cite{borjesson2012vgt, alegroth2013jautomate}, there is no significant difference between the tools and therefore does not affect the validity of the results.

\subsection{Phase 1: Interview study}
In phase 1, an interview study was performed at Siemens Medical, a company that develops life-critical medical journal systems for nurses and doctors.
The studied project was developed by a group of 15 developers and testers working according to an agile development process based on Scrum.
Verification of system requirement conformance was performed with unit testing, manual scenario-based and exploratory testing as well as manual acceptance testing with end users.
Seven months prior to the study the group had introduced VGT, with the tool JAutomate~\cite{alegroth2013jautomate}, into their test process in an attempt to lower test cost and raise quality.
JAutomate was introduced through development of test scripts as 1-to-1 mappings of existing manual test cases in the project.
At the time of the study, approximately 100 out of 500 manual test cases had been automated to be used for continuous integration.
However, because of development issues the testers had not been able to make the test suite run automatically from the build server.
The scripts were instead started manually and executed several times a week.

The interviews conducted at Siemens Medical aimed at eliciting the current state-of-practice at the company as well as the interviewees' experiences with VGT and the VGT tool JAutomate.
Three interviews were held with three interviewees using an interview guide consisting of 8 questions with sub-questions, 35 questions in total. 
The guide's first 4 questions gave insight into the company's context, questions 5 and 6 elicited the challenges the company had experienced that required the adoption of VGT and the final two questions elicited the interviewees' experiences with the technique.
The guide was constructed prior to the interviews and was reviewed internally in the research team prior to use.
Additionally, an analysis was made of the first interview's results to evaluate the suitability of the interview guide to give information for the study's purpose.
The guide was found suitable and no changes were therefore made for the latter interviews.
This conclusion was strengthened by the exploratory nature of the study, which implied the use of semi-structured interviews, which allowed the researchers could ask follow-up questions as required to get more detailed information.
Thus, some deviations were made in what follow-up questions were asked in the three interviews but the questions were kept as consistent as possible to ensure that the elicited information cold be triangulated between the respondents' answers.

The interviews were conducted with three testers at the company that had prior to their work with JAutomate worked with both manual scenario-based testing as well as other test automation techniques.
At the time of the study the testers had used VGT for several months, making them suitable to answer questions regarding the adoption, usage as well as maintenance of VGT scripts.
However, due to budget constraints, the interviewees were chosen through convenient sampling but included, by chance, the tester that had instigated the adoption as well as the test manager for the team.
The sample of interviewees was therefore considered suitable for the study.

All the interviews were recorded and then transcribed for further analysis where a simple coding scheme was used to extract answers to each interview question that were then stored in an excel sheet.
In addition to the interview questions' answers, other information of interest from the interviews, e.g. from follow-up questions, was also stored.
All the acquired information was then triangulated between interviewees as a measure of result validity.
Finally, the analyzed results were documented in a report that was submitted to the interviewees for validation after which no changes were made to the analyzed data, i.e. the authors' analysis was considered valid by the interviewees.

\subsection{Phase 2: Study Setting}\label{p2cs}
In phase 2, an empirical study was performed at Saab where a VGT test suite for an air-traffic management system, developed in the open-source VGT tool Sikuli~\cite{chang2010gui}, was maintained in several steps.

Saab is a developer of safety-critical software with roughly 80 employees in Sweden, split between two development sites, one in Gothenburg and one in V\"axj\"o.
The study was performed in V\"axj\"o where a reference system was made available for the research team for the study.
The company develops a set of products using both plan-driven and agile development processes for both domestic and international airports.
Most of the company's testing is performed through rigorous manual scenario-based system testing but also automated unit-testing in some projects.
The rigorous test process is required for the company's products to be compliant with the RTCA DO-278 quality assurance standard~\cite{kornecki2009certification}.

VGT has been sparsely used at the company through the use of the VGT suite that was created in previous work~\cite{alegrothvisual}.
The sparse use had however, at the time of the current study, degraded the test suite~\cite{berner2005observations} to a point where it was no longer executable on the SUT.
The VGT suite was initially developed as a 1-to-1 mapping of the manual test suite for an older version of the SUT, in the continuation of this work referred to as System X version 1.0.
System X was chosen in our previous work due to its complexity, airport runway and radar control, and size, in the order of 100k lines of code.
In addition, System X has a shallow graphical user interface (GUI), meaning that interaction with one GUI element of the GUI does not hinder the interaction with any other GUI element on the screen for instance by opening a blocking dialog window.
This GUI property is beneficial for GUI based testing since it reduces the script's logical complexity by, for instance, mitigating the need for synchronization between script and SUT required when opening menus or changing between GUI states to reach the expected output.

System X's VGT suite included 31 test cases, with an average of 306 lines of code per test script (Standard deviation 185) of Python code.
Each test case was divided into an average of eight test steps (standard deviation 5) that each contained interactions to set the system in the specific state for verification.
As such, the test steps could in another context be considered separate test cases and each test case a test suite due to their size and relative complexity.
We emphasize this property of the studied test suite since test case size and structure differs significantly in related work on GUI based testing.
For instance, in the work of Leotta et al.~\cite{leotta2013capture}, where the development and maintenance cost of written and recorded Selenium scripts were compared for six web systems, each test suite had an average of 32 test cases with a total average of 523 Selenese or 2390 Java lines of code per test suite.
To be compared to the approximately 9500 Python lines of code in the studied VGT test suite (not including support scripts) at Saab.
The studied VGT suite was developed in a custom test framework, also written in Python, which was created in previous work to support the creation of test suites that is not supported by the Sikuli tool. 
A detailed description of the test suite architecture and the framework can be found in~\cite{alegrothvisual}.

\begin{figure}[t!]
  \centering
  \includegraphics[scale=0.5]{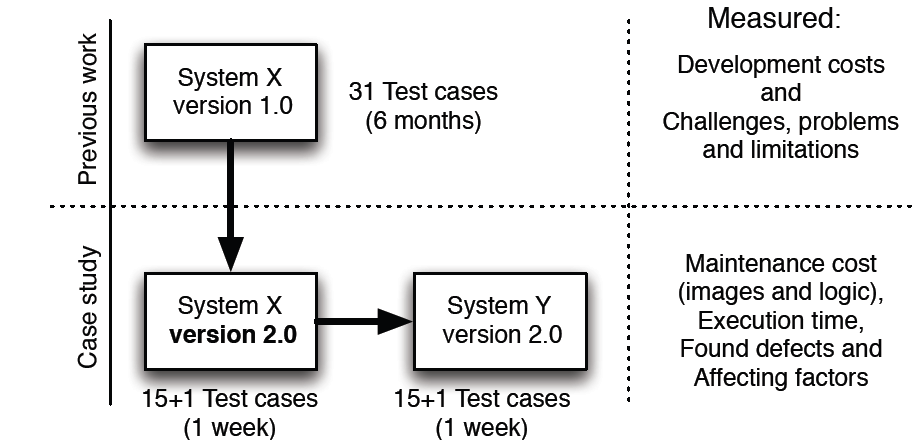}
  \caption{An illustration of the methodology used during the maintenance of the VGT test suite step (Step 2), shown in Figure\ref{meth}. The methodology started with maintenance in a big bang fashion between System X version 1.0 to 2.0 of the system. Second, System X version 2.0 was maintained for System Y. In both steps qualitative and quantitative metrics were collected, metrics listed to the right in the figure.}
  \label{method2}
\end{figure}

Phase 2 was divided into two steps, as shown in Figure \ref{meth}.
In the first step, the manual test specification used to develop the VGT suite for System X version 1.0 was analyzed and compared to the manual test specification of System X version 2.0.
In addition, version 1.0 was compared to another variant of the system, meaning another version of the system intended for another customer, we will refer to as System Y, version 2.0.
Hence, the study included three VGT suites and three systems, i.e. System X version 1.0, System X version 2.0 and System Y version 2.0, as shown in Figure \ref{method2}.
These three systems and test suites were included in the study since they allow the study of two different maintenance tasks; one smaller and one larger.

The analysis in step 1 gave insight into the required effort for the maintenance and let us estimate the feasibility of performing the study, as visualized in Figure \ref{method2}.
Step 2 then covered a period of two calendar weeks in which the actual VGT maintenance was done for both the smaller and larger maintenance tasks. The two weeks was the amount of time that Saab gave us access to the reference system and Saab personnel.

The analysis in step 1 was performed through manual inspection of each test step by comparing them for all textual test case descriptions between the different test specifications of the three systems.
Identified discrepancies were then evaluated to estimate the amount of effort that would be required to maintain/migrate each script.
The estimations were performed by a member of the research team with knowledge about both System X, the manual test specifications and the Sikuli testing tool.
The required maintenance effort was estimated on a scale from 1 to 10 where 1 was low effort and 10 was high effort.
No correctional actions were taken to mitigate estimation bias but the estimations were discussed with practitioners at Saab who reported them as realistic.
However, the lack of bias mitigation can have affected the estimations' validity, discussed further in Section \ref{disc}.

Based on the analysis results, 15 representative test cases were chosen to be maintained in the study. 
Representativeness was judged based on required effort to ensure that both test cases that required low and high effort were chosen.
In addition, the properties of individual test cases were taken into account, for instance if they included loops, branches, required interaction with animated and non-animated GUI components, required support of one or several airport simulators, etc.
As such, the 15 selected test cases were chosen to cover diverse aspects while still being possible to maintain in the allotted two-week period.
Other measures of diversity, e.g. code coverage, was not used due to time- and technical constraints, i.e. it was not possible to measure the code coverage of the manual tests in a cost-efficient way due to the SUT's complexity.
Whilst this lack of other metrics is considered a threat, it is evaluated as minor due to the common use of the manual test cases chosen for the study and the company's expert opinions of the representativeness of the test cases.

After the study, the estimated maintenance efforts for the chosen scripts were correlated against the recorded maintenance effort (measured as time) to test the hypothesis:
\begin{itemize}
\item $H_{01}$: It is possible to estimate the effort required to maintain a VGT script through analysis of the changes made to the manual test specification. 
\end{itemize}
Note that this hypothesis assumes that the VGT scripts are created as 1-to-1 mappings of the manual test cases, share test steps, interactions and/or at least have comparable test objectives.
Hence, that the automated test architecture is comparable to the scenario specified by the manual test case.

The second step of phase 2 was performed on site at Saab over a period of two calendar weeks (80 work-hours) through hands-on maintenance of the 15 representative test cases that were chosen in step 1.
Maintenance is in this context defined as, \emph{the practice of refactoring a test script to ensure its compliance with a new version of its manual test specification and/or to make it executable on a new version of the system under test.}

\begin{figure}[t!]
  \centering
  \includegraphics[scale=0.6]{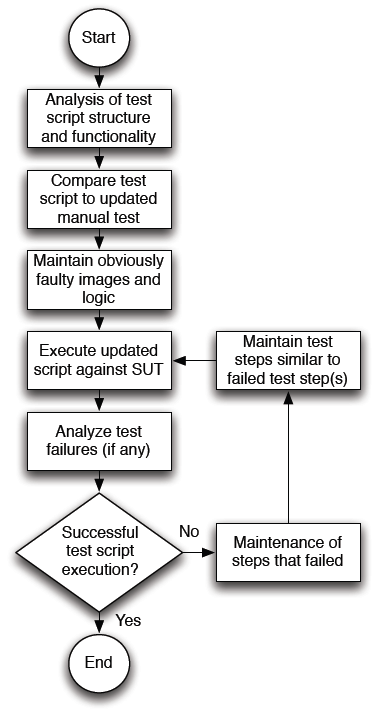}
  \caption{Visualization of the structured maintenance process that was used during the study.}
  \label{maintenance_process}
\end{figure}

The maintenance process used during the study followed a structured approach visualized in Figure \ref{maintenance_process}.
This process was developed for the study since it was observed during previous work that maintenance of a VGT script scenario is error-prone due to the dependencies between different test steps, i.e. test step X is required for test step X+1 to be successful.
In longer scripts it therefore becomes difficult for a tester/developer to keep track of all the scenario's steps and the order in which they are executed.
The proposed/used process mitigates this problem by breaking the maintenance effort down into smaller pieces that can be individually verified for correctness.
However, script verification can require multiple test runs which is tedious and therefore, as we will discuss later, scripts should be kept as short and linear as possible.

The VGT suite maintenance was performed, using pair programming, by one member of the research team and a resource from Saab with expert knowledge about the domain and System X. 
Pair programming was used because a secondary objective of the study was to further transfer VGT knowledge to Saab.
Activities that were performed during the maintenance included change of script logic (test scenarios) in order to test changed functionality of the SUT as well maintenance of GUI bitmaps.
Measures were not taken for change of independent lines of code or specific maintenance actions because these varied between different test scripts based on the intent of the script.
Instead, time/cost measurements were restricted to a script level of abstraction where the division of cost related to maintenance of logic was separated from the cost of maintaining GUI bitmaps.

The maintenance was performed in two parts, as visualized in Figure \ref{method2}.
Observant readers will notice that the figure states that there were 15+1 test cases in the test suite.
This additional test case was developed during the study based on domain knowledge, rather than the manual test specification, to evaluate if such test cases would be more or less costly to maintain.
The test case was written early during the maintenance process by the industrial practitioner that was part of the maintenance team and aimed to test each button of a controller for airport TAXI lighting.
Hence, lighting used by an airplane to find its way from the runway to its designated parking area or vice versa.

\subsection{Phase 2: Study Procedure}\label{p2cs2}
The first part of phase 2 began by maintaining the 15 chosen test cases from System X version 1.0 to System X version 2.0.
In addition, due to changes of System X's functionality and operational environment, substantial effort was required to maintain the VGT test framework itself.
Framework changes included refactoring of support scripts and methods, such as fine-tuning of visual toggling between subsystems, support for new simulators, etc.
In addition, the tests were refactored by replacing direct image paths in the scripts with variables that were stored in a single support script.
Thereby making the scripts more modular by placing all images, used in all test scripts, in a single script reused by all other scripts.
Because the support scripts were of equivalent functionality, complexity and behavior as the test scripts, all qualitative and quantitative metrics were recorded equivalently for all scripts regardless of type, with the exception of execution time.
The total set of maintained scripts were as such 21+1 scripts of which 15+1 were test cases and six were support-related.

In the second part of phase 2, System X was replaced by System Y which was another variant of System X for another airport with a GUI with minor functional and appearance differences.
The maintained VGT suite for System X version 2.0 was then maintained for System Y with the purpose of identifying indicative costs associated with regular maintenance of a VGT suite.
Hence, a context where the VGT suite had been used and maintained for frequent, or near frequent, integration due to the closer similarities between the different variants of the SUT.
The acquired information was used to test the hypothesis:
\begin{itemize}
\item $H_{02}$: There is no significant difference in cost to maintain the VGT suite from version 1.0 to version 2.0 of System X compared to migrating the VGT suite from System X version 2.0 to System Y version 2.0.
\end{itemize}
The result of this hypothesis evaluation provides support to the claim that the cost of frequent maintenance, with smaller deltas in terms of changed system functionality, is less costly than big-bang maintenance efforts~\cite{berner2005observations}.

The same metrics were collected in both parts of phase 2, results shown in Table \ref{script_data}.
More explicitly, the metrics that were measured were maintenance cost (time), time spent on maintaining script logic and images, execution time for running the resulting test case, number of found defects as well as qualitative observations, e.g. tool limitations and missing system functionality.

The maintenance cost metric was measured from the start of the maintenance of a new script until it could be executed successfully against System X/Y.
Hence, the measured time includes both the time spent on refactoring the script and the time required to verify that the script executed correctly against the SUT.
Cost was measured this way to mirror the effort required in a realistic context where test script maintenance has to be followed by verification of test script correctness.

However, the development costs of the VGT suite for System X, version 1.0, collected in our previous work~\cite{alegrothvisual}, only included the time spent on developing script code, not including execution.
As such, in order to be able to compare the maintenance costs with the development costs, the maintenance costs had to be transformed by removing the average number of test executions for verification during maintenance multiplied with the script execution time.
Hence, comparative maintenance time ($T_{X.maint.corrected}$) of a test case was evaluated according the formula,
\[
T_{X.maint.corrected} = T_{X.maint.measured} - (\bar{n}_{T.exe} * T_{X.exe})
\]
where $T_{X.maint.measured}$ is the true maintenance time including execution time for verification, $\bar{n}_{T.exe}$ is the observed number of average verification runs per test case (constant) and $T_{X.exe}$ is the execution time of the test case.
Note that $T_{X.maint.corrected}$ is used for all comparisons between development and maintenance cost in this work but that $T_{X.mait.measured}$ is used in favor of $T_{X.maint.corrected}$ whenever possible.
Our reason for limiting the use of $T_{X.maint.corrected}$ is because the number of reruns required to verify the correctness of a script during development or maintenance, $\bar{n}_{T.exe}$, fluctuated between scripts.
As such, the constant $\bar{n}_{T.exe}$, equal to the average of required reruns of all scripts, introduces a marginal error in the comparative maintenance value, $T_{X.maint.corrected}$.
Since this error presents a threat to validity, the reader will be informed when $T_{X.maint.corrected}$ has been used in favor of $T_{X.maint.measured}$ and what impact its use might have had on the presented result.
Furthermore, since the support scripts were invoked by test scripts during runtime, their execution time could not be measured, as shown in Table \ref{script_data}.
Therefore, in order to calculate $T_{X.maint.measured}$ for said scripts, $T_{X.exe}$ was set to the average execution time of the scripts in respective VGT suites.
Thus introducing an additional error for these scripts.

The quantitative information was then analyzed statistically to test the previously stated hypothesis, $H_{02}$, and the following null hypotheses.
\begin{itemize}
\item $H_{03}$: There is no significant difference between the cost to maintain images and logic in a script.
\item $H_{04}$: There is no significant difference between the cost to develop and the cost to maintain a VGT suite. 
\end{itemize}
Hypothesis $H_{03}$ was analyzed to evaluate the maintenance costs on a finer level of granularity and is perceived important for GUI based test techniques that use scenario-based scripts that interact with GUI components.
GUI components that can be defined as bitmaps as in VGT or as constructs based on the properties of the GUI components as in second generation GUI based test tools.
In turn, hypothesis $H_{04}$ aims to investigate if the costs of maintaining the automated scripts are less than the development cost.
Since the development costs of the test cases has been shown to be feasible but significant~\cite{alegroth2013vgt, alegrothvisual}, acceptance of $H_{04}$ would infer that automated testing is not feasible.

During the second phase of the study, triangulation was performed between the interview results from Phase 1 and the observations made in Phase 2. 
For instance if the challenges mentioned by the industrial practitioners at Siemens could be observed, in situ, during maintenance of the VGT scripts at Saab.
Thus allowing triangulation of results also across the development companies that were part of the study, enhancing the external validity of the study's conclusions.
Hence, after the two phases, the collected qualitative results were compared to generalize the results and draw the final conclusions.

The collected results were analyzed using formal statistical tools, e.g. correlation and data comparison, to support the qualitative results.
Two different tools were chosen, correlation analysis to evaluate $H_{01}$ and comparative hypothesis testing analysis for $H_{02}$ to $H_{04}$.
Due to the empirical nature of the results they were first analyzed for normality to establish if parametric tests could be used or if non-parametric tests were required.
For the comparative hypothesis testing, the parametric Student T-test and the non-parametric Wilcoxon test were considered since they evaluate the null hypothesis ($H_{0}$) that there is no statistical significant difference between two groups.
The collected quantitative metrics were used as input to the tools to evaluate the hypotheses.

In addition, the quantitative data was visualized in the theoretical cost model presented previous work~\cite{alegrothvisual, berner2005observations} to model the total cost of VGT maintenance in comparison to the cost of manual testing performed at Saab.
Manual testing that in Saab's context represents seven (7) percent of the total time spent in a software development project.
However, as stated, the costs of verification and validation (V\&V) in general software engineering practice lies in the bound of 20-50 percent of the total time.
The model therefore also includes a plot from a hypothetical but realistic context where the lower bound of 20 percent is spent on V\&V.
This context is modeled in order to visualize its effects on the time required to reach positive return on investment of test automation.
However, it is important to note that decisions to automate are not taken only from a cost perspective; there might be other benefits of automation than its effects on cost.
This study is however delimited only to cost and future work thereby includes analysis of other factors such as defect finding ability, impact of frequent failure localization, etc., on the total development cost of a project.

\section{Results and Analysis} \label{res} 
This section will present the results that were acquired in the embedded study, starting with the quantitative results from Phase 2 that will then be discussed in relation with the qualitative results from Phase 1 and 2.
The reason for presenting the results in this order is because the qualitative results help explain the quantitative results, e.g. what factors affected the maintenance effort.
Also, note that we in the following section present maintenance cost as time.

\subsection{Quantitative results}\label{qres}
The quantitative results collected during the second phase of the study have been summarized in Table \ref{script_data}.
The table columns present, in order:
\begin{enumerate}
\item the test case number (\#),
\item the test case company specific unique ID,
\item the test script's original development time, 
\item the test script's required maintenance time (minutes) from System X version 1.0 to 2.0,
\item the test script's required maintenance time (minutes) from System X version 2.0 to System Y version 2.0,
\item the distribution of maintenance of logic and images for System X version 1.0 to 2.0,
\item the distribution of maintenance of logic and images for System X version 2.0 to System Y version 2.0,
\item the script's execution time on System X version 2.0, and
\item the script's execution time on System Y version 2.0.
\end{enumerate}

Further, the rows of the table has been divided into three blocks, where the first block (rows 1 to 16) of rows present the data acquired during maintenance of test scripts.
In the second block (rows 17-22) the acquired data for the maintenance of support scripts are presented.
Finally, the last block (rows 23-24) summarizes the mean values and standard deviation of each column.
We once again stress that the development costs presented in this table does not take the time required to verify script correctness, through execution, into account, whilst the maintenance costs do.

\begin{table} 
\begin{center}
  \begin{tabular}[b]{| p{0.3cm} | p{1.2cm} | p{1.1cm} | p{1.3cm} | p{1.3cm} | p{1.35cm} | p{1.35cm} | p{1cm} | p{1cm} |}
  \hline
   \textbf{\#} & \textbf{Test script} & \textbf{Orig. dev. time (min)}	 & \textbf{Maint. 1.0X-2.0X (min)}  & \textbf{Maint. 2.0X-2.0Y (min)} &  \textbf{(logic/ img.) 1.0X-2.0X (min)}    &  \textbf{(logic/ img.) 2.0X-2.0Y (min)}   & \textbf{Exe. 2.0 X (min)} & \textbf{Exe. 2.0 Y (min)} \\ \hline
	1&	t0016	& 130	&	100	&	30	& 20/80		& 0/30		& 3.2		& 3.25		\\ \hline
	2&	t0017	& 110	&	100	&	35	& 70/30		& 0/35		& 3.167	& 2.816		\\ \hline
	3&	t0018	& 265	&	70	&	35	& 35/35		& 0/35		& 3.417	& 3.05		\\ \hline
	4&	t0019	& 250	&	230	&	20	& 138/92		& 0/20		& 3.083	& 3.067		\\ \hline
	5&	t0014	& 225	&	195	&	15	& 136.5/ 58.5	& 7.8/5.2		& 0.95	& 4.633		\\ \hline
	6&	t0003	& 245	&	120	&	10	& 0/120		& 0/10		& 0.75	& 0.783		\\ \hline
	7&	t0024	& 641	&	320	&	65	& 224/96		& 19.5/ 45.5	& 10.5	& 11.783		\\ \hline
	8&	t0005	& 705	&	215	&	10	& 107.5/ 107.5	& 0/10		& 1.85	& 2.067		\\ \hline
	9&	t0023	& 145	&	315	&	150	& 220.5/ 94.5	& 45/105		& 9.233	& 19.4		\\ \hline
	10&	t0007	& 370	&	10	&	5	& 9/1			& 0/0			& 1.783	& 1.8			\\ \hline
	11&	tS0001	& 20		&	20	&	10	& 12/8		& 0/10		& 1.416	& 1.416		\\ \hline
	12&	t0026	& 155	&	20	&	5	& 4/16		& 0/0			& 0.333	& 0.45		\\ \hline
	13&	t0009	& 40		&	50	&	10	& 10/40		& 1/9			& 2.05	& 1.5			\\ \hline
	14&	t0008	& 180	&	35	&	5	& 10.5/ 24.5	& 0/0			& 4.483	& 4.5			\\ \hline
	15&	t0041	& 140	&	35	&	20	& 10.5/ 24.5	& 0/20		& 4.617	& 4.517		\\ \hline
	16&	t0037	& 140	&	120	&	20	& 84/36		& 0/20		& 7.183	& 7.617		\\ \hline \hline
	17&	vncS.	& 415	&	250	&	50	& 175/75		& 35/15		& N/A	& N/A		\\ \hline
	18&	SimS.	& 30		&	40	&	0	& 28/12		& 0/0			& N/A	& N/A		\\ \hline
	19&	sysS.	& 370	&	15	&	0	& 6/9			& 0/0			& N/A	& N/A		\\ \hline
	20&	winS.	& 600	&	15	&	0	& 15/0		& 0/0			& N/A	& N/A		\\ \hline
	21&	SimS. t2	& 70		&	105	&	45	& 84/21		& 40.5/4.5		& N/A	& N/A		\\ \hline
	22&	windS.	& 300	&	110	&	0	& 55/55		& 0/0			& N/A	& N/A		\\ \hline \hline
	23&	\textbf{Ave.}	& \textbf{252}	&	\textbf{110}	&	\textbf{23}	& \textbf{66.1/ 47.1}		& \textbf{6.763/ 17.009}		& \textbf{3.626}	& \textbf{4.541}		\\ \hline
	24&	\textbf{Std. Dev.}	& \textbf{195}	&	\textbf{105}	&	\textbf{37}	& \textbf{71.634/ 37.485}	& \textbf{14.353/ 23.838}	& \textbf{2.989}	& \textbf{4.866}		\\ \hline
    \end{tabular}
    \caption{Summary of the collected metrics for the scripts that were maintained during the study. The test cases have been listed in the chronological order they were maintained in (denoted tXXXX), whilst support scripts have been listed out of order in the bottom of the list.}
    \label{script_data}
\end{center}
\end{table}

Analysis of the presented results in table \ref{script_data} show that the maintenance costs for transitioning the VGT suite from System X version 1.0 to System X version 2.0 was higher (mean 110) than the maintenance costs associated with transitioning the VGT suite from System X version 2.0 to System Y version 2.0 (mean 23).
Worth noting is that the standard deviation in the first case is almost as high as the average and in the second case almost twice as large.
This large deviation is attributed to many factors, such as script size, number of images, number of used simulators, etc.
In addition, the deviation is also influenced by the sample being right-skewed above 0.
We will return to these factors and their impact in Section \ref{factors}.

Correlation analysis was used to evaluate $H_{01}$, i.e. to evaluate if the required maintenance effort can be estimated based on changes to the system's specification.
The analysis showed that the correlation between estimated and actual effort between System X 1.0 and System X 2.0 was 0.165 and for System X 2.0 to System Y 2.0 was 0.072.
Hence, we reject our hypothesis, $H_{01}$, that it is possible to estimate the required maintenance effort based on the differences in the manual test specifications for VGT scripts based on said specifications.
The reason for the poor estimates relate to the many factors that affect the maintenance cost, which will be presented in Section \ref{factors}, but which could not be foreseen prior to the study.
However, the 15 chosen test cases were still considered a representative sample, supported by the distribution of measured maintenance costs, script functionality, etc.

Further, $H_{02}$ stated that the cost of maintaining the VGT suite for System X version 2.0 would not be statistically significantly different from the cost of maintaining the test suite for System Y.
Hypothesis testing with the non-parametric Wilcoxon test resulted in a p-value result of 7.735e-05.
Hence, we must reject the null hypothesis, $H_{02}$, at a 0.05 significance level.
Analysis of the collected data, shown in Table \ref{script_data}, shows that the transition between System X version 2.0 and System Y is lower than the cost for System X 1.0 to System X 2.0.
Consequently, since System X version 2.0 and System Y were more similar than System X version 1.0 to System X version 2.0, judged by experts at Saab, this result implies that less maintenance effort is required between more frequent releases.
This conclusion also supports previous research into automated testing that stipulate that maintenance should be performed with high frequency to mitigate maintenance costs, for instance, caused by test case degradation~\cite{berner2005observations}.

$H_{03}$ aimed to test if there is any statistically significant difference between the cost of maintaining images and logic in a VGT script.
The p-value result from the Wilcoxon test for the maintenance from System X 1.0 to System X 2.0 was 0.8972.
Hence, we can not reject the null hypothesis, $H_{03}$, that there was no statistical significance difference between the maintenance required of images and logic at the 0.05 confidence level.
This result shows that degraded VGT test suites require maintenance of images to the degree that it is equivalent to the maintenance of script logic.
However, for the maintenance of System X version 2.0 to System Y the p-value result of the Wilcoxon test was 0.01439.
Hence, we must reject the null hypothesis, $H_{03}$, that the maintenance cost of logic is not significantly different from the cost of maintaining images.
Analysis of the average maintenance costs of images and logic for the two maintenance efforts shows that more effort was spent on updating logic in the maintenance from System X version 1.0 to version 2.0 (66.1 minutes for logic and 47.1 for images) compared to System X version 2.0 to System Y (6.7 minutes for logic and 17 for images).
This observation helps explain why the maintenance costs were lower for the transition between System X 2.0 to System Y since an empirical observation was made that images were easier to update than test script logic.
These observations also support that maintenance should be performed frequently to lower the need to maintain both images and logic at the same time since it is assumed that GUI graphics are changed more often than the underlying functionality of safety-critical software.

Finally, $H_{04}$ aimed to test if the cost of maintaining the VGT suites was significantly different from the development cost of the suite.
The p-value result of the Wilcoxon test for the maintenance costs for System X version 1.0 to 2.0 compared to the development cost was 0.001868 and for System X version 2.0 to System Y 7.075e-08\footnote{Tested using the approximative value $T_{X.maint.corrected}$.}.
Consequently we must reject the null hypothesis, $H_{04}$, in both cases showing that there is statistical significant difference between the development costs and the maintenance costs.
Hence, this result shows that the maintenance costs are lower than the development costs.
Furthermore, analysis of the costs, shown in Table \ref{script_data}, shows that the average costs are lower than the development cost, in both cases, even when the verification time is included in the maintenance cost.
However, the transition between System X version 1.0 and 2.0 was 61.5 percent of the development cost and between System X version 2.0 and System Y 17.5 percent of the development cost of the VGT suite.
As such, the maintenance costs of a VGT suite are still significant.
We will return to a more detailed discussion about the impact of these results in Sections \ref{costmod} and \ref{disc}.

VGT is first and foremost a testing technique and as such, in addition to cost, its effectiveness must be judged based on its defect finding ability.
During the study, eight defects were found, either during execution or maintenance of the VGT scripts.
These defects were found through test cases that were divided evenly across the test suite, i.e. no specific test case found more defects.
In addition, these defects were of different type, such as GUI defects, e.g. buttons not working as expected, simulator defects, e.g. misalignment between simulator and system behavior, manual test documentation discrepancies, e.g. faulty test steps, etc.
As such providing support to previous work on VGT regarding the technique's defect finding ability~\cite{alegroth2013vgt,alegrothvisual}.
In addition, the spectrum of identified defects indicate that VGT is suitable for system level testing rather than testing of the GUI itself.

\begin{figure}[t!]
  \centering
  \includegraphics[scale=0.5]{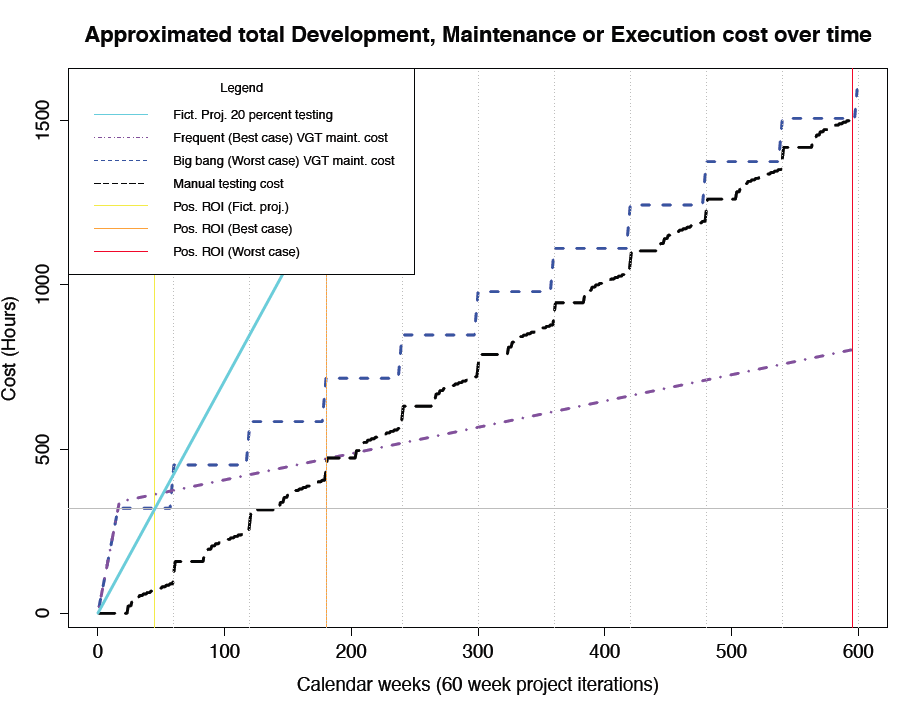}
  \caption{Plot showing the total cost of development and maintenance in the best and worst case based on the measurements acquired in the study. The graph also shows the cost of manual system testing at Saab and the point where the VGT testing reaches positive return on investment compared to manual testing. 
 }
  \label{AutoVsMan}
\end{figure}

\subsubsection{Modeling the cost}\label{costmod}
In our previous work we presented a theoretical return on investment (ROI) model for the development and maintenance of automated tests~\cite{alegrothvisual}.
This model was inspired by previous work~\cite{berner2005observations} and depicts that the linearly increasing cost of manual testing will surpass the combined cost of VGT script development and continuous maintenance after a period of time.
Thus resulting in a overall lower cost of testing if the tests are automated.
However, prior to this work the model was only theoretical.

In Figure \ref{AutoVsMan}, the actual, numerical results from phase 2 have been visualized in the proposed model, which shows how the total cost of script development and maintenance grows over time (calendar weeks) for a test suite with 70 test cases, equal to the number of manual scenario based test cases available for the system.
Four cost plots have been included in the model with the colors cyan, blue, black and purple.
First, the blue line (Short dashes) shows the total cost of development and maintenance if maintenance is performed only once every development iteration (60 weeks) in a big-bang fashion where all test scripts are maintained at once.
Second, the purple line (Long dashes with dots) shows the cost of development and regular maintenance of the VGT suite.
The line was calculated using the average best case maintenance cost identified in the study with the assumption that at most two scripts would require maintenance each week.
This assumption was verified with three practitioners at Saab, at different occasions, who all stated that two tests would be an upper bound due to the slow development rate of their system.
Third, the black line (Long dashes) shows the cost of manual testing of System X/System Y at Saab and is based on data from previous development iterations of the system, which is calculated to roughly seven (7) percent of the total project time.
The cost of manual testing was developed together with, and verified by, three different practitioners at Saab that had been and/or were responsible for the quality assurance of System X/Y and depicts/assumes the following,
\begin{itemize}
\item Each development iteration includes two complete regression tests of the manual test specification/suite, one at the start and one at the end.
\item Each development iteration includes delta tests based on the manual test specification/suite.
\item It assumes that there is no down period between development iterations where the system is not being developed.
\end{itemize}
These assumptions, and previous approximations, adds error to the model but based on the verification made by Saab's experts and validity analysis, based on the collected metrics from the study, it is perceived as a valid approximation.
Detailed and exact cost data is almost never reliably found in actual, industrial development projects and these types of estimations should be expected even if a detailed time logging system is in place.

Finally, the cyan line (Solid) represents the costs of testing in a fictional project where 20 percent of the time is spent on testing.
Unlike in the Saab case, the fictional project's testing is the total cost of all test activities, i.e. unit testing, system testing, acceptance testing, etc.

In addition to the costs, the graph also contains three vertical lines that depict the points in time when the automation provides positive return on investment (ROI) compared to manual testing.
First, the yellow line (Furthest to the left) Figure \ref{AutoVsMan} shows when VGT adoption would provide positive ROI in the fictional project, which is after roughly 45 weeks (approximately 11 months).
Second, the orange line (Middle) shows when the adoption of VGT provides positive ROI compared to the manual test costs at Saab if the test suite is maintained regularly, which is after 180 weeks (3.5 years) or 3 iterations of System X/Y.
Third, the red line (Furthest to the right) shows when positive ROI would be achieved at Saab if the maintenance is performed in a big-bang fashion, which is after 532 weeks (Approximately 10 years) or roughly 9 iterations.

Note that Figure \ref{AutoVsMan} only presents ROI as a measure of cost.
However, there are other factors that affect ROI as well, such as the frequency of feedback on system quality to the developers and defect finding ability of the different techniques.
As such, the figure only presents a partial ROI model that does not consider the overall cost benefits for the project such as shortened defect analysis time, raised software quality, etc.
These factors are required to acquire a holistic model on the ROI of automated testing but are not further explored in this work.
Further, we have chosen to represent cost for positive return on investment as a measure of time despite other measures could have been used, e.g. number of executions.
However, since we are measuring maintenance costs versus the cost of manual testing, we believe that this would be the most fair comparison.

\subsection{Qualitative results}
This section will present the qualitative results that were acquired in phase 1 and 2 of the study and discuss them in relation to the previously presented quantitative results. We decided to present the results in this order since the qualitative results can more easily be understood when read in light of the quantitative data of phase 2, as reported above.

\subsubsection{Phase 1: Interview results}
In phase 1 an interview study was performed with three interviewees at one division within Siemens Medical that seven months prior to the study introduced VGT with the tool JAutomate.
Prior to the introduction of JAutomate all system and acceptance testing had been performed manually by testers or the company's end customers, i.e. nurses and doctors.
These tests were performed during a dedicated test phase.
\emph{``There is something here at Siemens called a system test phase, the system is tested as a whole when all of the development is completed.''}
In addition to the manual testing the company also used automated unit and performance testing for lower levels of system abstraction testing.
\emph{``For the system, we had unit tests for as much as possible to test the code''.}
The main reasons for the introduction of JAutomate were to lower test cost and to create a test harness for continuous system testing.
\emph{``We just got three machines to run automatic builds on where we are now installing JAutomate''.}
\emph{``We will run this at night, JAutomate, after an installation or new build. Then we don't have to run the boring tests that otherwise are manual''.}
Thus mitigating the need for frequent, costly and tedious manual system testing~\cite{grechanik2009maintaining, cadar2011symbolic}.

The tested system was composed of a server and a client application where the server side was covered by a rigorous unit test suite.
However, the client side had proved to be difficult to test with unit tests due to its GUI focus, which left a need for another automated test approach.
An attempt had been made to cover the GUI testing with the tool Coded test UI but had been unsuccessful because the tool was too costly to work with.
\emph{``It took roughly the same time to automate a couple of hundred test cases (with JAutomate) as it took to start building tests in Coded UI (after creating a Coded UI framework)''.}

The introduction of JAutomate, including the transition from manual tests to automated JAutomate test cases, took roughly 4 calendar months, following a three month evaluation period.
During this time, 100 out of the available 500 manual test cases had been automated and were reported to have a total execution time of roughly four to five hours.
Whenever a JAutomate test case was created from a manual test case the company had chosen to remove the manual test from the test plan.
\emph{``Yes, we did (remove the manual test cases), but we also wrote many new tests when manual tests didn't exist. Then we only created a JAutomate test''.}
However, the interviewees still stated that JAutomate should not be considered a replacement for manual testing but rather a complement.
\emph{``Complement, because manual tests test other things than a strict script does''.}
Therefore they still ran the manual test cases during release testing of the system but much less frequently than prior to the adoption of VGT.

When asked about the return on investment (ROI) of the JAutomate adoption, the interviewees were uncertain but they perceived it to have been beneficial for the developed system.
\emph{``Hard to say, but we could cut the system test part a great deal which is the important part in this case''.}

Furthermore, the interviewees stated that they found the technique beneficial because the VGT tests are more similar to human testing and because the technique is very strict in terms of execution, which contributes to its defect-finding-ability.
\emph{``It has the benefit that it repeats it (tests) in the same way. It is powerful that way...a human can not always see the whole picture and react on a discrepancy in that picture. For instance that something is fussy or a discrepant''.}
In addition, the interviewees found VGT with JAutomate easy to learn and perceived that others could learn it easily given that the user was not technically awkward.
To support this claim, the interviewees stated that the company had performed experiments with the system's end users, e.g. nurses, that were asked to write test scripts with JAutomate.
However, due to their lack of computer knowledge the experiment had not been successful and a conclusion was drawn that the quality gains in comparison to cost of teaching the end users to use JAutomate were less than having the end users execute the tests manually.
\emph{``Maybe they (end user) can learn,..., everyone can learn but they have a longer way to go''.}
This result partially contradicts the result from our previous work where it was found that technically unsophisticated users could write suitable VGT scripts after one hour of tutoring~\cite{borjesson2012vgt}. However, these results were based on VGT with Sikuli and not JAutomate.
Further research is therefore required to evaluate the learnability of the different techniques.

In addition, the interviewees reported that the main problem with JAutomate was the tool's robustness.
\emph{`It's the robustness, to get everything to work, to get it to go green (pass) 100 times out of a 100. Also to check that it works here, it is pretty slow''.}
A problem they reported to be primarily related to timing issues rather than the tool's image recognition.
\emph{``You have to spend a large amount of time to handle the programs updates and wait for things and similar, which causes things that worked three weeks ago to stop working.''}
Timing is a common problem for GUI testing tools, especially for web-based systems where network latency has to be taken into account.
The problem originates in how to determine when the tested system has reached a stable state such that it is ready to receive new input.
Other issues that were presented were that the script development was slow in comparison to the development of new functionality in the system that made it difficult to get system coverage, tester education to create good tests, structuring the test cases, etc.
\emph{``We (the development team) changed quite a lot in a day, but we still feel that their tests worked. It has to be the toughest task there is. To create UI tests to something under development, agile development. We really made changes to everything''.}

One of the interviewees also reported that the maintenance could take up to 60 percent of the time spent on JAutomate testing each week, whilst the retaining 40 percent consisted of transition or development of new test cases.
Hence, more effort was spent each week on maintaining existing test cases than to write new ones.
However, when asked if the amount of required maintenance was feasible, two of the interviewees said yes and the third said no.
\emph{``Yes, it (maintenance) is (feasible) since you can get it to run, hopefully, during nightly runs and such, which makes you more secure in the quality of the product and development.''}
Furthermore, when asked if the interviewees trusted the tool, one interviewee said no, one said to 95 percent and the last said yes.
However, when asked if the interviewees would recommend the tool to other project teams at Siemens, all of them said that they would.
In addition all of the interviewees stated that they found the tool very fun to work with and that it was beneficial to the company.
However, when asked about JAutomate's worst features the interviewees stated that the robustness, as reported, is low, the tool is slow, that there are logistic problems with large test suites and that the symbiosis with Microsoft testing software could be better.

In summary, the interview results showed that VGT with JAutomate requires a lot of time to be spent on test script maintenance, i.e. up towards 60 percent of the time spent each week, which included execution time of the test scripts.
Furthermore the tool suffers from robustness problems in terms of timing and technical issues when it comes to integration in a build process for continuous integration.
However, the tool and the technique are still considered beneficial, valuable, fun and mostly feasible by the practitioners.
As such, the interviewees still regarded the benefits of the tool and the technique to outweigh the drawbacks.
Hence supporting the results from our previous work with Sikuli at Saab in J\"arf\"alla~\cite{alegroth2013vgt}.
The acquired results on positive ROI after roughly seven months of working with the technique also support the results from Saab and previous work~\cite{alegrothvisual} since the maintenance was still perceived as feasible after this time.

\subsubsection{Phase 2: Observations}\label{obs}
Previous research into VGT has reported that verifying script correctness is frustrating, costly and tedious~\cite{alegroth2013vgt,alegrothvisual}.
Especially for long test cases since verification requires the script developer to observe the script execution and for each failed execution update the script and then rerun it.
This observation was also made for maintenance of the scripts, which infers that test scripts should be kept as short and modular as possible.
How long a test case should optimally be was not analyzed during this work and is therefore a subject of future research but it is perceived that it is dependent on context.

Furthermore, it was observed that lengthy test cases were also harder to understand and more complex because of lack of overview of the script's behavior.
Higher complexity was especially observed in test scripts that contained loops and/or several execution branches.
This observation lead us to the conclusion that not only should test cases be kept short but also as linear as possible, a practice that also make the scripts more readable by other developers.
Test cases that tested multiple features in one script were also observed as more complex to analyze and therefore more costly to maintain.
Hence, the number of features verified per test script should be kept as few and similar, i.e. coupled, as contextually possible.
These observations lead us to the conclusion that 1-to-1 mapped test cases, despite the benefit of allowing test script verification against the manual test specification, are not necessarily the most suitable automation scheme. 
Especially for scenario-based system and acceptance tests that generally verify many requirements at once, instead 1-to-1 mappings between test cases and requirements are perceived as a better alternative.

To facilitate these best practices, more requirements are put on the test suite architecture to support modular test case design and test partitioning.
Reuse of modular scripts is perceived beneficial to creating several test scripts with incrementally longer test chains since changes to the behavior of the SUT will require all affected scripts to be maintained.
Furthermore, script code reuse, as opposed to linking reusable scripts together, makes it tempting to copy and paste script code.
Copy and paste was observed as an error-prone practice during the study since the scripts are dependent both on the GUI images as well as the synchronization between script and SUT behavior and should therefore be avoided.

Further it was observed that the correctness of the meta-level/support scripts had direct influence on the success and cost of maintenance due to the reuse of these scripts in the test scripts.
Thus, placing more stringent requirements on the robustness of the support scripts.
However, script robustness assurance requires additional code that in addition to raising cost also raises script execution time~\cite{alegrothvisual}.
Thus, supporting the need for frequent maintenance to ensure that the support scripts up to date.

\subsubsection{Phase 2: Factors that affect the maintenance of VGT scripts}\label{factors}
Based on observations from the second phase of the study, triangulated with the interviews from phase 1 and literature~\cite{rafi2012benefits, wagner2006model, karhu2009empirical, liu2000platform, berner2005observations, fewster1999software, sjosten2006costs, leotta2013capture}, we found 13 factors that impact test script maintenance, summarized in Table \ref{factors}.
Impact was classified into four degrees, which are low, average, high and total.
Low impact means that the presence of this factor will lengthen the maintenance time with a maximum of a few minutes, average impact factors will add several minutes up to one hour, high impact factors more than one hour and finally total impact will prevent the test case to be maintained.
Each of the factors have been discussed from the context of VGT but generalized to other automated test practices where applicable or perceived applicable.

\begin{table}
\begin{center}
  \begin{tabular}[b]{| p{0.5cm} | p{2.3cm} | p{7cm} | p{1.5cm} |}
  \hline
   \textbf{Nr} & \textbf{Factor}	 & \textbf{Description}  & \textbf{Impact} \\ \hline
	1	& Nr. of images		& Each image that requires maintenance raises cost. 	& Low	\\ \hline
	2	& Knowledge/ Experience		& A VGT script expert can maintain a script quicker than a novice.	& Low	\\ \hline\hline
	3	& Mindset 	& VGT scripting requires a sequential mindset that differs from traditional programming.  & Average	\\ \hline
	4	& Variable names and script logic	& Common to traditional programming. Better code structure improves readability.	& Average	\\ \hline
	5	& Test case similarity	& Difference between the system and/or test specification compared to old version.	& Average	\\ \hline
	6	& Meta level script	& Meta level script functionality adds complexity and is costly to maintain. 	& Average	\\ \hline\hline
	7	& Test case length	& Long test cases are more complex and less readable. 	& High	\\ \hline
	8	& Loops and flows in the test case	& Loops and alternative test flows affect the test scripts' complexity and readability. 	& High	\\ \hline
	9	& Simulator(s)	& Simulators can require special interaction code.	& High	\\ \hline\hline
	10	& System functionality	& Missing system functionality can hinder test script maintenance.	& Total	\\ \hline
	11	& Simulator support	& Lacking simulator support can hinder test script maintenance.	& Total	\\ \hline
	12	& Defects		& Defects in the tested system case hinder test script maintenance.	& Total	\\ \hline
	13	& VGT tool	& Limitations in the test tool can hinder test script maintenance.	& Total	\\ \hline
	    \end{tabular}
    \caption{Summary of the main factors that were identified that affect the required maintenance effort of a VGT suite. The impact of each factor has been categorized from Low to Total. \emph{Low} impact is defined as a required extra cost to develop a script of a few minutes maximum. \emph{Average} is defined as an increased cost (time) of a few minutes up to an hour. \emph{High} is defined as an increased cost (in time) more than an hour. \emph{Total} means that this factor can completely prevent the script from being maintained.} 
    \label{factors}
\end{center}
\end{table}

\emph{Number of images.} 
Image maintenance is associated with low cost and is therefore considered low impact.
However, the cost increases with the number of images in the script and can therefore become substantial for long scripts since broken images are only found during test script execution.
A good practice is therefore to always check if a broken image is reused more than once in the script and/or replace duplicate images with variables if supported by the VGT tool.
This factor implies that script maintenance is split into test logic (test scenario) and interaction components (images).
Consequently, this factor is perceived common to second generation tools where images are replaced with interaction components that are instead GUI component properties.

\emph{Knowledge/Experience.}
VGT has average/high learnability, as shown in this and previous research~\cite{borjesson2012vgt}, due to the high level of interaction that the scripts work with.
The knowledge and experience of the user therefore has low impact on script maintenance.
However, it is a good practice to have a domain expert perform VGT development and/or maintenance to ease analysis and implementation of domain specific knowledge in the scripts.
Knowledge/experience is considered to have a larger impact on automated test techniques that operate on a lower level of system abstraction since they require more technical and domain knowledge.

\emph{Mindset.}
VGT script development/maintenance requires a sequential mindset that differs from traditional programming since the scripted scenarios need to be synchronized with the timing of the tested system.
This factor is considered to have average impact since it is sometimes difficult even for a VGT expert to anticipate when synchronization points are required to align script execution with the tested system's state transitions, which can cause frustration.
This factor has been presented in previous work both on VGT and automated testing in general~\cite{alegrothvisual, berner2005observations, karhu2009empirical}, and to the best of our knowledge there is no mitigation practice.
Further, synchronization has been identified as a common problem for automated GUI level test techniques and this factor is therefore considered common to other GUI based test techniques as well.
 
\emph{Variable names and script logic.}
Common to traditional software development, VGT script complexity impairs script code readability, reusability and maintainability~\cite{berner2005observations} and therefore has average impact on maintenance costs.
To mitigate the impact, a good practice is to up front define a clear and consistent test script architecture and define naming conventions for variables and methods.
This factor is perceived to be general to all automated testing tools that use more advanced script logic.

\emph{Test case similarity.}
Previous work has shown that test suites degrade if not frequently used and maintained~\cite{berner2005observations}, resulting in higher maintenance costs due to increased failure analysis complexity.
Degradation is caused by development/maintenance of the tested system and/or changes to the system requirements or manual test cases if the VGT scripts are developed in a 1-to-1 mapping fashion.
The impact of this factor is considered average but also general to all automated test techniques. 
A practice to mitigate the impact of this factor is to frequently maintain the test scripts.

\begin{figure}[t!]
  \centering
  \includegraphics[scale=0.3]{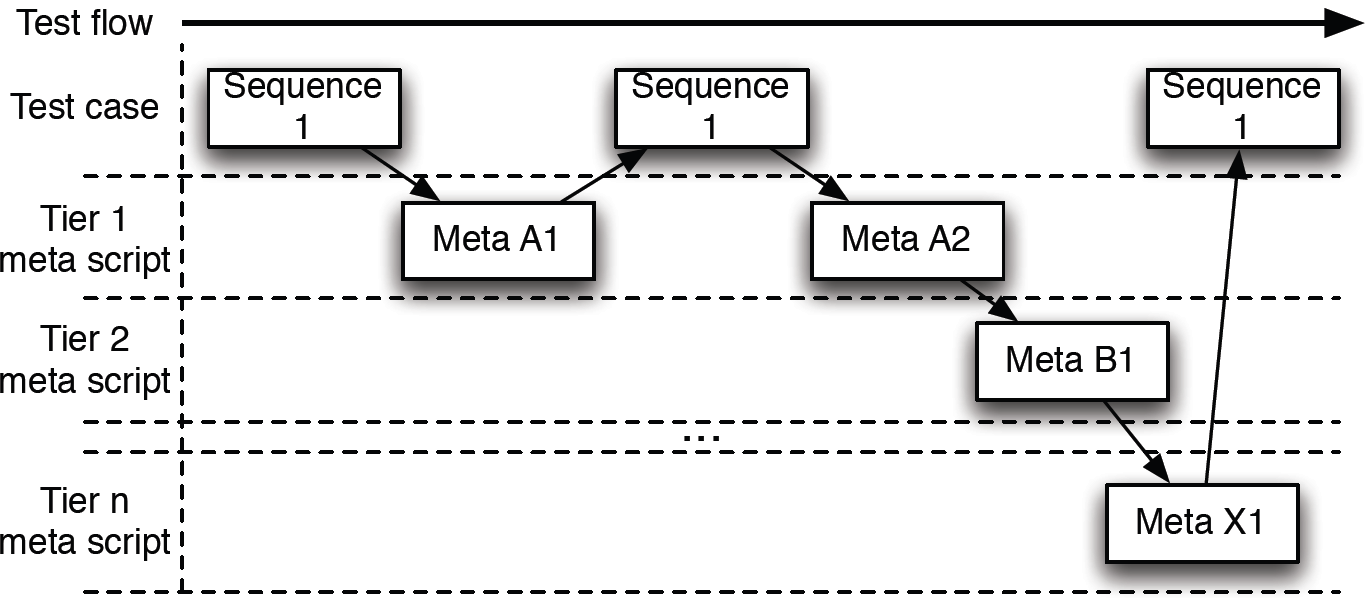}
  \caption{Abstract model of the GUI interaction sequences within a test script, visualizing the complexity of understanding test cases including meta-level scripts.}
  \label{vgt_sequence}
\end{figure}

\emph{Meta level script.}
Meta level scripts facilitate test script operation by performing interactions with the tested system's environment that are not part of the test scenario, e.g. start and modify simulators or modify the test system's configuration.
Meta level interactions is common in system level tests which put more stringent requirements on the meta level scripts' robustness.
Their common use in a test case also lowers test scenario readability, as illustrated in Figure \ref{vgt_sequence}.
As such, this factor is associated with average impact on maintenance but can be mitigated through up front investment on meta script robustness, e.g. through implementation of additional failure mitigation and exception handling in core scripts.
The factor is also considered common to other test techniques and tools that support modular test design where meta level scripts are generally reused to set the system state for certain assertions.

\emph{Test case length.}
Long test scripts are less readable and more complex to maintain because of lack of overview.
In addition, longer scripts take longer to execute and verify, which can cause frustration in the case of heavy script degradation.
A good practice is therefore to keep test scripts short.
Alternatively the scripts should have a modular architecture that allows for subsets of the script to be maintained individually from the rest of the script.
Because long scripts are common and not always possible to avoid, the impact of this factor is considered high and also general to other automated test techniques, especially GUI based test automation.

\emph{Loops and flows in the test case.}
Loops and branching test flows should be avoided in VGT scripts because they lower readability and make script failure analysis more complex.
As such, a good practice is to keep scripts as linear as possible and break loops and/or branches into individual scripts.
Because alternative script flows cannot always be avoided the impact of this factor is considered high and also general to other automated test techniques that support more advanced script logic.

\emph{Simulator(s).}
The purpose of simulators is to emulate software or hardware that will be part of the system's operational environment.
Simulators are often crude in terms of GUI design and can be developed in other programming languages than the tested system itself.
As such, the impact of having simulators is high for VGT. 
However, for other automated test techniques it is total due to restrictions of the tools' applicability for certain programming languages and distributed systems, e.g. second generation tools.
Hence, this factor is general to automated testing but with varying impact that depends on the the other techniques'/tools' capabilities.
The factor can be mitigated for other techniques if the simulators can be operated through simulator APIs.

\emph{System functionality.}
VGT scripts are used for system regression testing and therefore require the tested system to have reached a certain level of implementation to be applicable.
Missing or changed functionality can therefore lead to unmaintainable or partially maintained test scripts.
As such, the impact of this factor can be total and considered general for techniques that support automated system and acceptance testing.
Previous research has shown that the lack of system functionality will limit how much of the test case can be implemented~\cite{alegrothvisual}.
This factor is considered low for lower level automated test techniques such as unit testing.

\emph{Simulator support.}
Added or changed functionality in the tested system can cause simulators to stop working completely or partially.
Partially working simulators can lead to test script failure and inability, or only partial ability, to maintain said test scripts until the simulator itself has been maintained.
Thus, the impact of this factor can be total but can be mitigated through frequent maintenance of the tested system's environment.
This factor is considered general to all test automation techniques as well as manual testing.

\emph{Defects.}
Defects in the system limits the ability to maintain a test script beyond the test step that finds said defect because further interaction with the tested system would be in a defective system state that would never appear in practice.
Thus, this factor can have total impact on the script maintenance and a good practice is therefore to not execute and only maintain the affected test scripts once the defect in the SUT has been resolved.
This factor is considered general for both manual and automated testing.

\emph{VGT tool.}
Different VGT tools have different functionality which make them more or less suitable in different contexts.
For instance, only some VGT tools have support to verify that the system can play sound, others have script recording functionality or support for testing of distributed systems.
Failure to pick the right tool for the right context can cause this factor to have total impact on both the development and maintenance of test cases.
Previous work has compared different VGT tools against each other~\cite{borjesson2012vgt, alegroth2013jautomate} but how to pick the most suitable VGT tool for a certain context is still a subject of future work.
This factor is also perceived to be common to all automated test tools since they all have different capabilities.

\section{Discussion} \label{disc}  
The main conclusion of this work is that automated testing, represented by VGT, can provide positive ROI over time when applied in industrial practice despite requiring considerable maintenance.
In particular, the results show that the maintenance costs associated with automated test scripts are lower than the development cost of the scripts, shown with statistical significance, independent of if worst or best case maintenance practices are used.
Worst case data was acquired empirically through measurement of the costs of maintaining a heavily degraded test suite between two versions of a system at Saab.
This was followed by the migration of the maintained test suite to a similar variant of the studied system to acquire approximate but valid data of best case maintenance costs of a VGT suite.

Average maintenance cost in the best case was found to be 23 minutes per test script and 110 minutes in the worst case, whilst the cost of manual execution of a test case was 29 minutes.
Plotting the acquired results over time, in a theoretical cost model defined in previous work~\cite{alegrothvisual, berner2005observations}, Figure \ref{AutoVsMan} shows that test automation would provide positive ROI at Saab in 180 weeks in the best case and 532 weeks in the worst case compared to manual testing.
However, even 180 weeks represent a considerable, long-term, investment to reach positive ROI and the maintenance costs of the acquired test suite would still be significant, upwards of 60 percent of the time spent on test automation each week as reported by Siemens.
These results are however placed in Saab's context where approximately seven (7) percent of the total project time is spent on manual testing.
Therefore, Figure \ref{AutoVsMan} also includes a plot of the verification and validation (V\&V) costs in a fictional context where 20 percent of the project time is spent on V\&V, which shows that positive ROI is reached in 45 weeks.
However, 20 percent is still a lower bound, according to research, of the time spent on V\&V in practice that generally spans between 20-50 percent of the total development time of a project~\cite{ellims2006economics, hailpern2002software,ericson1997tim}. 
These related results imply that ROI in another context could be reached even faster.
Consequently, the time to positive ROI of test automation is directly dependent on the time spent on V\&V at the company prior to automation.

Further, observations from the study show that maintenance of automated tests are dependent on several factors, of which thirteen (13) were identified and presented in this work.
These factors include technical factors, e.g. test case length, organizational factors, e.g. the tester's knowledge and experience, environmental factors, e.g. simulator support, etc.
Whilst these factors were identified for VGT it is perceived that they are common to other automated test techniques as well, but especially other GUI based test techniques since they, as discussed in previous work, have many commonalities~\cite{alegrothvisual}.
However, future work is required to verify this statement.

The implications of the presented results and observations are that there are many aspects to consider when adopting test automation in practice.
First, test maintenance costs are significant and continuous due to the identified need for frequent test case maintenance to mitigate cost.
As such, automated testing will necessarily not lower the time spent on V\&V in a project but perceivably lowers the overall project development time by providing frequent feedback regarding the quality of the SUT.
This feedback allows the developers to identify defects quicker and thereby mitigate synergy effects between defects that is perceived to lower defect root cause analysis time.
Thus, the primary benefit of test automation is raised software quality rather than lowered test related costs.
Further, the factors identified in this work that affect test maintenance imply that the technical aspects of the automation, e.g. the test architecture, are key to ensure maintainability of the test scripts.
This conclusion in turn implies that best practice of traditional software development should be applied when creating an automated test suite, for instance the test architecture should be modular, there should be code standards for scripting, interfaces to the SUT's environment must be well defined, etc.
However, it is not enough to consider the technical aspects, one also needs to consider organizational and human factors such as the knowledge and experience of the tester (tool and domain knowledge), that the tester has a sequential mindset to fulfill the need to synchronize the script execution with the SUT execution, etc.
Further support for these factors' impact on automated testing has been presented in previous work~\cite{karhu2009empirical, berner2005observations}.
However, since the identified and presented factors are complementary to factors presented in previous work it is perceivable that more factors exist that were not observed during this study.
As such, further work is required to identify more factors, what the impact of the different factors are in relation to each other and ways to mitigate the negative effects of said factors. 

More specifically, this work also implies that VGT can be applied in industrial practice and provide positive ROI compared to manual testing.
Thus partially bridging the gap for a technique for automated system and acceptance testing~\cite{borjesson2012vgt,alegroth2013vgt,alegrothvisual}.
It should be noted that the cost data presented in this work does not take the perceived software quality gains provided by more frequent testing and faster feedback to the developers compared to manual regression testing into account.
As reported in Section \ref{qres}, eight defects were identified during the study either during maintenance or execution of the test suite.
The identified defects provide support to previous work regarding VGT's defect finding ability~\cite{alegroth2013vgt,alegrothvisual}.
In addition, the found defects indicate that additional cost savings can be made with VGT through faster defect identification and identification of defects that cannot be found feasibly through manual testing, e.g. defects that appear seldom during runtime.
All of the identified defects were reported to the company and has since the study been maintained.
However, explicitly how the technique's defect finding ability affects the time to positive ROI is still a subject of future research.

Finally, this work provides a general contribution to the body of knowledge on test automation.
Previous work has contributed with empirical results~\cite{rafi2012benefits, wagner2006model, karhu2009empirical, liu2000platform, berner2005observations, fewster1999software, sjosten2006costs, leotta2013capture} but more is required, especially for GUI based test automation techniques, e.g. VGT, which currently only has limited cost support from industrial practice~\cite{alegroth2013vgt}.
Additional empirical support is required to gain a more holistic view on the maintenance costs associated with general automated testing in industry, which requires more studies, in more companies and with different techniques and/or tools.

\subsection{Threats to validity}
One threat to the external validity of this study is that the quantitative data was only acquired for one system and with one VGT tool.
Furthermore, the study only evaluated frequent maintenance of the suite in a cross-sectional manner rather than longitudinal.
As such, even though the results support that more frequent maintenance is associated with lower cost than big-bang maintenance, more work is required to analyze the costs associated with maintenance in a longitudinal perspective.
The results from Siemens indicate that such maintenance is feasible, but more quantitative data is required.
In addition, both systems are safety critical, which could imply more stringent constraints and requirements on the software development process and software quality than for general software.
However, as the VGT tools observe any system, regardless of complexity, as a black-box the results provided by this study are considered general to most GUI based software.

Another threat, related to the internal validity of the results, is that only 21+1 data points could be acquired during the study.
However, as stated, the test cases in this work are more comprehensive than test cases presented in related work on GUI test automation~\cite{leotta2013capture}.
Analysis of the test cases indicate that, in terms of size and complexity, each of the test cases could according to previous research be classified as test suites.
The average number of test steps per script was seven (7), meaning that the results of this work could be generalized to a context where 147 test cases had been maintained rather than 147 test steps as classified in this work.

Yet another internal validity threat was that only three people were interviewed at Siemens.
However, conclusions have only been made on data that could be triangulated from all three sources.
Furthermore, since the interviewees were the leading developers on the VGT implementation project it is considered that their statements are credible.

Finally, the estimated maintenance costs used to test hypothesis $H_{01}$ were acquired by only one person in the research team.
As such, the estimates could have been affected by researcher bias.
However, the researcher performing the estimations had expert knowledge about the manual tests, the VGT tool, scripting, etc., making this person highly suitable to perform the estimation.
Regardless, we can not rule out that the researcher may have overestimated or underestimated the costs of the maintenance.
Thus, future work is required to verify our results.

One threat to the construct validity of the study is that the two maintenance efforts were performed on two variants of the same system rather than versions.
However, analysis of the system's specifications, supported by statements by the system's developers, indicated that the differences between System X and System Y are few in terms of GUI appearance and functionality, which supports its use for measurement of frequent maintenance.
This observation does however imply that further work is required to verify our results on several consecutive versions of the same system or application.

Another threat was that the VGT maintenance costs had to be measured in a different manner, including test execution time for verification, than the development costs reported in previous work~\cite{alegrothvisual}.
However, since it was possible to calculate reasonable estimations of the time spent on maintenance programming using the equation in section \ref{p2cs}, this threat is considered low.
To verify the output from the equation, the calculated values were discussed and validated with the leading researcher from the previous study.
In addition, the execution time of the scripts is small compared to development and maintenance time of the scripts and as such the observed maintenance costs are still comparable to the development costs without augmentation.
Additionally, the augmented data was only used to verify the hypothesis that the development time was not significantly different from the maintenance time, whilst all other statistical analysis was done with the true observations.
The reader has also been notified whenever the augmented data has been used in the manuscript.

These threats can partially be explained by the study being performed in industry under both cost and time constraints that limited the amount of empirical work that could be performed.

\section{Conclusions} \label{conc}
The main conclusion of this work is that automated testing, represented by VGT, can provide positive return on investment (ROI) compared to manual testing in industrial practice.
This conclusion was drawn based on data from two different companies, Siemens and Saab.
The results show, with significance, that the development costs are greater than the maintenance costs and that the costs of frequent script maintenance are lower than big-bang maintenance.
However, the costs of test script maintenance are still substantial, identified at Siemens to constitute upwards of 60 percent of the time spent on test automation each week.
A cost that can be compared to the academically reported costs that companies spend on verification and validation (V\&V), which range between 20-50 percent of the total time spent in a project~\cite{ellims2006economics, hailpern2002software,ericson1997tim}.
Thus, the magnitude of required time spent on maintenance is equal to the costs of V\&V in general, which infers that test automation may not lower time spent on V\&V.
However, automated testing can raise the trust in the quality of a system due to more frequent quality feedback that could lower overall project development time.
As such, the study shows that the time to reach positive ROI is dependent on the amount of V\&V performed by the company prior to automation.

Additionally, thirteen (13) qualitative factors were observed during the study that affect the maintenance costs of VGT scripts but also automated testing in general, e.g. developer knowledge/experience, developer mindset, test case length and test case linearity.
The identified factors infer that maintenance of test scripts, but also development, depend on an intricate balance of technical an non-technical aspects in order to reach qualitative tests.
Whilst these factors provide a general contribution to the body of knowledge on automated testing, further research is required to identify complementary factors and to measure their impact.

Additionally, this work provides explicit support for the use of VGT in practice by showing that positive ROI can be reached and by providing further support to previous work regarding VGT's defect finding ability.
Eight (8) defects were identified during the study, spread among the SUT and its operational environment, e.g. the SUT's simulators.
In combination with related work~\cite{borjesson2012vgt,alegroth2013vgt,alegrothvisual}, these results show that VGT is a feasible complement to other manual and automated test techniques in practice to facilitate automated system and acceptance testing.

\bibliographystyle{IEEEtran}
\bibliography{bib}

\begin{thebibliography}{10}
\providecommand{\url}[1]{#1}
\csname url@samestyle\endcsname
\providecommand{\newblock}{\relax}
\providecommand{\bibinfo}[2]{#2}
\providecommand{\BIBentrySTDinterwordspacing}{\spaceskip=0pt\relax}
\providecommand{\BIBentryALTinterwordstretchfactor}{4}
\providecommand{\BIBentryALTinterwordspacing}{\spaceskip=\fontdimen2\font plus
\BIBentryALTinterwordstretchfactor\fontdimen3\font minus
  \fontdimen4\font\relax}
\providecommand{\BIBforeignlanguage}[2]{{%
\expandafter\ifx\csname l@#1\endcsname\relax
\typeout{** WARNING: IEEEtran.bst: No hyphenation pattern has been}%
\typeout{** loaded for the language `#1'. Using the pattern for}%
\typeout{** the default language instead.}%
\else
\language=\csname l@#1\endcsname
\fi
#2}}
\providecommand{\BIBdecl}{\relax}
\BIBdecl

\bibitem{ellims2006economics}
M.~Ellims, J.~Bridges, and D.~C. Ince, ``The economics of unit testing,''
  \emph{Empirical Software Engineering}, vol.~11, no.~1, pp. 5--31, 2006.

\bibitem{hailpern2002software}
B.~Hailpern and P.~Santhanam, ``Software debugging, testing, and
  verification,'' \emph{IBM Systems Journal}, vol.~41, no.~1, pp. 4--12, 2002.

\bibitem{ericson1997tim}
T.~Ericson, A.~Subotic, and S.~Ursing, ``Tim - a test improvement model,''
  \emph{Software Testing Verification and Reliability}, vol.~7, no.~4, pp.
  229--246, 1997.

\bibitem{olsson2012climbing}
H.~H. Olsson, H.~Alahyari, and J.~Bosch, ``Climbing the" stairway to heaven"--a
  mulitiple-case study exploring barriers in the transition from agile
  development towards continuous deployment of software,'' in \emph{Software
  Engineering and Advanced Applications (SEAA), 2012 38th EUROMICRO Conference
  on}.\hskip 1em plus 0.5em minus 0.4em\relax IEEE, 2012, pp. 392--399.

\bibitem{olan2003unit}
M.~Olan, ``{Unit testing: test early, test often},'' \emph{Journal of Computing
  Sciences in Colleges}, vol.~19, no.~2, pp. 319--328, 2003.

\bibitem{gamma1999junit}
E.~Gamma and K.~Beck, ``{JUnit: A cook's tour},'' \emph{Java Report}, vol.~4,
  no.~5, pp. 27--38, 1999.

\bibitem{holmes2006automating}
A.~Holmes and M.~Kellogg, ``Automating functional tests using selenium,''
  \emph{|}, pp. 270--275, 2006.

\bibitem{hackner2008test}
D.~R. Hackner and A.~M. Memon, ``{Test case generator for GUITAR},'' in
  \emph{Companion of the 30th international conference on Software
  engineering}.\hskip 1em plus 0.5em minus 0.4em\relax ACM, 2008, pp. 959--960.

\bibitem{vizulisself}
V.~Vizulis and E.~Diebelis, ``{Self-Testing Approach and Testing Tools},''
  \emph{Datorzin{\=a}tne un inform{\=a}cijas tehnolo{\`g}ijas}, p.~27, 2012.

\bibitem{borjesson2012vgt}
E.~B{\"o}rjesson and R.~Feldt, ``{Automated System Testing using Visual GUI
  Testing Tools: A Comparative Study in Industry},'' in \emph{Software Testing,
  Verification and Validation (ICST), 2012 IEEE Fifth International Conference
  on}.\hskip 1em plus 0.5em minus 0.4em\relax IEEE, 2012, pp. 350--359.

\bibitem{alegroth2013vgt}
E.~Alegroth, R.~Feldt, and H.~Olsson, ``{Transitioning Manual System Test
  Suites to Automated Testing: An Industrial Case Study},'' in \emph{Software
  Testing, Verification and Validation (ICST), 2013 IEEE Sixth International
  Conference on}.\hskip 1em plus 0.5em minus 0.4em\relax IEEE, 2013, pp.
  56--65.

\bibitem{alegrothvisual}
E.~Al{\'e}groth, R.~Feldt, and L.~Ryrholm, ``Visual gui testing in practice:
  challenges, problems and limitations,'' \emph{Empirical Software
  Engineering}, pp. 1--51, 2014.

\bibitem{leotta2014visual}
M.~Leotta, D.~Clerissi, F.~Ricca, and P.~Tonella, ``Visual vs. dom-based web
  locators: An empirical study,'' in \emph{Web Engineering}, ser. Lecture Notes
  in Computer Science.\hskip 1em plus 0.5em minus 0.4em\relax Springer, 2014,
  vol. 8541, pp. 322--340.

\bibitem{rafi2012benefits}
D.~M. Rafi, K.~R.~K. Moses, K.~Petersen, and M.~Mantyla, ``Benefits and
  limitations of automated software testing: Systematic literature review and
  practitioner survey,'' in \emph{Automation of Software Test (AST), 2012 7th
  International Workshop on}.\hskip 1em plus 0.5em minus 0.4em\relax IEEE,
  2012, pp. 36--42.

\bibitem{wagner2006model}
S.~Wagner, ``A model and sensitivity analysis of the quality economics of
  defect-detection techniques,'' in \emph{Proceedings of the 2006 international
  symposium on Software testing and analysis}.\hskip 1em plus 0.5em minus
  0.4em\relax ACM, 2006, pp. 73--84.

\bibitem{karhu2009empirical}
K.~Karhu, T.~Repo, O.~Taipale, and K.~Smolander, ``Empirical observations on
  software testing automation,'' in \emph{Software Testing Verification and
  Validation, 2009. ICST'09. International Conference on}.\hskip 1em plus 0.5em
  minus 0.4em\relax IEEE, 2009, pp. 201--209.

\bibitem{liu2000platform}
C.~Liu, ``Platform-independent and tool-neutral test descriptions for automated
  software testing,'' in \emph{Proceedings of the 22nd international conference
  on Software engineering}.\hskip 1em plus 0.5em minus 0.4em\relax ACM, 2000,
  pp. 713--715.

\bibitem{berner2005observations}
S.~Berner, R.~Weber, and R.~Keller, ``{Observations and lessons learned from
  automated testing},'' in \emph{Proceedings of the 27th international
  conference on Software engineering}.\hskip 1em plus 0.5em minus 0.4em\relax
  ACM, 2005, pp. 571--579.

\bibitem{fewster1999software}
M.~Fewster and D.~Graham, \emph{Software test automation: effective use of test
  execution tools}.\hskip 1em plus 0.5em minus 0.4em\relax ACM
  Press/Addison-Wesley Publishing Co., 1999.

\bibitem{sjosten2006costs}
E.~Sj{\"o}sten-Andersson and L.~Pareto, ``{Costs and Benefits of
  Structure-aware Capture/Replay tools},'' \emph{SERPS’06}, p.~3, 2006.

\bibitem{leotta2013capture}
M.~Leotta, D.~Clerissi, F.~Ricca, and P.~Tonella, ``Capture-replay vs.
  programmable web testing: An empirical assessment during test case
  evolution,'' in \emph{Reverse Engineering (WCRE), 2013 20th Working
  Conference on}.\hskip 1em plus 0.5em minus 0.4em\relax IEEE, 2013, pp.
  272--281.

\bibitem{runeson2009guidelines}
P.~Runeson and M.~H{\"o}st, ``{Guidelines for conducting and reporting case
  study research in software engineering},'' \emph{Empirical Software
  Engineering}, vol.~14, no.~2, pp. 131--164, 2009.

\bibitem{grechanik2009maintaining}
M.~Grechanik, Q.~Xie, and C.~Fu, ``{Maintaining and evolving {GUI}-directed
  test scripts},'' in \emph{Software Engineering, 2009. ICSE 2009. IEEE 31st
  International Conference on}.\hskip 1em plus 0.5em minus 0.4em\relax IEEE,
  2009, pp. 408--418.

\bibitem{grechanik2009creating}
------, ``{Creating {GUI} testing tools using accessibility technologies},'' in
  \emph{Software Testing, Verification and Validation Workshops, 2009.
  ICSTW'09. International Conference on}.\hskip 1em plus 0.5em minus
  0.4em\relax IEEE, 2009, pp. 243--250.

\bibitem{finsterwalder2001automating}
M.~Finsterwalder, ``{Automating acceptance tests for {GUI} applications in an
  extreme programming environment},'' in \emph{Proceedings of the 2nd
  International Conference on eXtreme Programming and Flexible Processes in
  Software Engineering}.\hskip 1em plus 0.5em minus 0.4em\relax Citeseer, 2001,
  pp. 114--117.

\bibitem{leitner2007reconciling}
A.~Leitner, I.~Ciupa, B.~Meyer, and M.~Howard, ``{Reconciling manual and
  automated testing: The autotest experience},'' in \emph{System Sciences,
  2007. HICSS 2007. 40th Annual Hawaii International Conference on}.\hskip 1em
  plus 0.5em minus 0.4em\relax IEEE, 2007, pp. 261a--261a.

\bibitem{memon2002gui}
A.~Memon, ``{GUI testing: Pitfalls and process},'' \emph{IEEE Computer},
  vol.~35, no.~8, pp. 87--88, 2002.

\bibitem{dustin1999automated}
E.~Dustin, J.~Rashka, and J.~Paul, \emph{{Automated software testing:
  introduction, management, and performance}}.\hskip 1em plus 0.5em minus
  0.4em\relax Addison-Wesley Professional, 1999.

\bibitem{cheon2006simple}
Y.~Cheon and G.~Leavens, ``{A simple and practical approach to unit testing:
  The JML and JUnit way},'' \emph{ECOOP 2002, Object-Oriented Programming}, pp.
  1789--1901, 2006.

\bibitem{autorev2012}
D.~Rafi, K.~Moses, K.~Petersen, and M.~Mantyla, ``{Benefits and limitations of
  automated software testing: Systematic literature review and practitioner
  survey},'' in \emph{{Automation of Software Test (AST), 2012 7th
  International Workshop on}}, june 2012, pp. 36 --42.

\bibitem{gutierrez2006generation}
J.~J. Guti{\'e}rrez, M.~J. Escalona, M.~Mej{\'\i}as, and J.~Torres,
  ``{Generation of test cases from functional requirements. A survey},'' in
  \emph{4{\c{s}} Workshop on System Testing and Validation}, 2006.

\bibitem{horowitz1993g}
E.~Horowitz and Z.~Singhera, ``{Graphical user interface testing},''
  \emph{Technical eport Us C-C S-93-5}, vol.~4, no.~8, 1993.

\bibitem{zaraket2012guicop}
F.~Zaraket, W.~Masri, M.~Adam, D.~Hammoud, R.~Hamzeh, R.~Farhat, E.~Khamissi,
  and J.~Noujaim, ``{GUICOP: Specification-Based GUI Testing},'' in
  \emph{Software Testing, Verification and Validation (ICST), 2012 IEEE Fifth
  International Conference on}.\hskip 1em plus 0.5em minus 0.4em\relax IEEE,
  2012, pp. 747--751.

\bibitem{afzalexperiment}
W.~Afzal, A.~N. Ghazi, J.~Itkonen, R.~Torkar, A.~Andrews, and K.~Bhatti, ``An
  experiment on the effectiveness and efficiency of exploratory testing.''

\bibitem{itkonen2007defect}
J.~Itkonen, M.~V. Mantyla, and C.~Lassenius, ``Defect detection efficiency:
  Test case based vs. exploratory testing,'' in \emph{Empirical Software
  Engineering and Measurement, 2007. ESEM 2007. First International Symposium
  on}.\hskip 1em plus 0.5em minus 0.4em\relax IEEE, 2007, pp. 61--70.

\bibitem{Itkonen2005}
J.~Itkonen and K.~Rautiainen, ``{Exploratory testing: a multiple case study},''
  in \emph{Empirical Software Engineering, 2005. 2005 International Symposium
  on}, nov. 2005, p. 10 pp.

\bibitem{adamoliautomated}
A.~Adamoli, D.~Zaparanuks, M.~Jovic, and M.~Hauswirth, ``{Automated GUI
  performance testing},'' \emph{Software Quality Journal}, pp. 1--39, 2011.

\bibitem{andersson2004video}
J.~Andersson and G.~Bache, ``{The video store revisited yet again: Adventures
  in {GUI} acceptance testing},'' \emph{Extreme Programming and Agile Processes
  in Software Engineering}, pp. 1--10, 2004.

\bibitem{gao2015}
Z.~Gao, Y.~Liang, M.~B. Cohen, A.~Memon, and W.~Z., ``Making system user
  interactive tests repeatable: When and what should we control?'' in \emph{The
  Proceedings of the 37th International Conference on Software Engineering
  (ICSE2015)}.\hskip 1em plus 0.5em minus 0.4em\relax IEEE, 2015.

\bibitem{vos2015testar}
T.~E. Vos, P.~M. Kruse, N.~Condori-Fern{\'a}ndez, S.~Bauersfeld, and
  J.~Wegener, ``Testar: Tool support for test automation at the user interface
  level,'' \emph{International Journal of Information System Modeling and
  Design (IJISMD)}, vol.~6, no.~3, pp. 46--83, 2015.

\bibitem{yeh2009sikuli}
T.~Yeh, T.~Chang, and R.~Miller, ``{Sikuli: using {GUI} screenshots for search
  and automation},'' in \emph{Proceedings of the 22nd annual ACM symposium on
  User interface software and technology}.\hskip 1em plus 0.5em minus
  0.4em\relax ACM, 2009, pp. 183--192.

\bibitem{alegroth2013jautomate}
E.~Alegroth, M.~Nass, and H.~Olsson, ``{JAutomate: A Tool for System-and
  Acceptance-test Automation},'' in \emph{Software Testing, Verification and
  Validation (ICST), 2013 IEEE Sixth International Conference on}.\hskip 1em
  plus 0.5em minus 0.4em\relax IEEE, 2013, pp. 439--446.

\bibitem{NguyenASE2013}
B.~N. Nguyen, B.~Robbins, I.~Banerjee, and A.~Memon, ``Guitar: an innovative
  tool for automated testing of gui-driven software,'' \emph{Automated Software
  Engineering}, vol.~21, no.~1, pp. 65--105, 2014.

\bibitem{chang2010gui}
T.~Chang, T.~Yeh, and R.~Miller, ``{GUI} testing using computer vision,'' in
  \emph{Proceedings of the 28th international conference on Human factors in
  computing systems}.\hskip 1em plus 0.5em minus 0.4em\relax ACM, 2010, pp.
  1535--1544.

\bibitem{kornecki2009certification}
A.~Kornecki and J.~Zalewski, ``Certification of software for real-time
  safety-critical systems: state of the art,'' \emph{Innovations in Systems and
  Software Engineering}, vol.~5, no.~2, pp. 149--161, 2009.

\bibitem{cadar2011symbolic}
C.~Cadar, P.~Godefroid, S.~Khurshid, C.~S. Pasareanu, K.~Sen, N.~Tillmann, and
  W.~Visser, ``{Symbolic execution for software testing in practice:
  preliminary assessment},'' in \emph{Software Engineering (ICSE), 2011 33rd
  International Conference on}.\hskip 1em plus 0.5em minus 0.4em\relax IEEE,
  2011, pp. 1066--1071.

\end{thebibliography}

\end{document}